      [1999/12/01 v1.4c Il Nuovo Cimento]
\documentclass[preprint]{cimento}

\usepackage{cite}
\usepackage{cancel}
\usepackage{amsmath}
\usepackage{graphicx}
\newcommand{\eq}[1]{Eq.~(\ref{#1})}
\newcommand{\ps}{p\hspace{-0.44em}/\hspace{0.06em}}
\newcommand{\Msusy}{M_{\textrm{SUSY}}}

\newcommand{\gev}{\, \mathrm{GeV}}
\newcommand{\tev}{\, \mathrm{TeV}}
\newcommand{\bbs}{\ensuremath{B_s\!-\!\overline{B}_s\,}}

\title{$B$ Decays and Lepton Flavour (Universality) Violation}
\author{A.~Crivellin\from{ins:x}}
\instlist{\inst{ins:x} CERN Theory Division, CH-1211 Geneva 23, Switzerland}
\PACSes{11.30.Hv,12.60.Cn,12.60.Fr,13.25.Hw,14.65.Fy,14.80.Da,14.80.Ly}
\begin{document}
\maketitle
\begin{abstract}
LHCb found hints for physics beyond the standard model in $B\to K^*\mu^+\mu^-$, $B\to K^*\mu^+\mu^-/B\to K^*e^+e^-$ and $B_s\to\phi\mu^+\mu^-$. In addition, the BABAR results for $B\to D^{(*)}\tau\nu$ and the CMS excess in $h\to\tau^\pm\mu^\mp$ also point towards lepton flavour (universality) violating new physics. While $B\to D^{(*)}\tau\nu$ and $h\to\tau^\pm\mu^\mp$ can be naturally explained by an extended Higgs sector, the probably most promising explanation for the $b\to s\mu\mu$ anomalies is a $Z'$ boson. Furthermore, combining a 2HDM with a gauged $L_\mu-L_\tau$ symmetry allows for explaining the $b\to s\mu^+\mu^-$ anomalies and $h\to\tau^\pm\mu^\mp$ simultaneously, with interesting correlations to $\tau\to3\mu$. In the light of these deviations from the SM we also discuss the possibilities of observing lepton flavour violating $B$ decays (e.g. $B\to K^{(*)}\tau^\pm\mu^\mp$ and $B_s\to\tau^\pm\mu^\mp$) in $Z^\prime$ models.
\end{abstract}

\section{Introduction}

With the discovery of the Brout--Englert--Higgs boson~\cite{Aad:2012tfa} the LHC completed the standard model (SM) of particle physics. While no new particles have been discovered at the LHC so far, some 'hints' for indirect effect of new physics (NP) in the flavor sector appeared: $B\to K^* \mu^+\mu^-$, $B_s\to\phi\mu^+\mu^-$, $R(K)=B\to K \mu^+\mu^-/B\to K e^+e^-$, $B\to D^{(*)}\tau\nu$ and $h\to\mu\tau$ deviate from the SM predictions by $2-3\,\sigma$, each. While for the first three anomalies only quark flavour must be violated by NP, the last three processes require the violation on lepton flavour (universality).

\subsection{Experimental hints for NP in the flavour sector}


LHCb reported deviations from the SM predictions~\cite{Egede:2008uy} (mainly in an angular observable called $P_5^\prime$~\cite{Descotes-Genon:2013vna}) in $B\to K^* \mu^+\mu^-$~\cite{Aaij:2013qta,LHCb:2015dla} with a significance of $2$--$3\,\sigma$ depending on the assumptions of hadronic uncertainties~\cite{Descotes-Genon:2014uoa,Altmannshofer:2014rta,Jager:2014rwa}. Also in the decay $B_s\to\phi\mu^+\mu^-$ \cite{Aaij:2013aln} LHCb uncovered differences compared to the SM prediction from lattice QCD \cite{Horgan:2013pva,Horgan:2015vla} of $3.1\,\sigma$ \cite{Altmannshofer:2014rta}. Furthermore, LHCb~\cite{Aaij:2014ora} found indications for the violation of lepton flavour universality in
\begin{equation}
	R(K)={\rm Br}[B\to K \mu^+\mu^-]/{{\rm Br}[B\to K e^+e^-]}=0.745^{+0.090}_{-0.074}\pm 0.036\,,
\end{equation}
in the range $1\,{\rm GeV^2}<q^2<6\,{\rm GeV^2}$ which disagrees with the theoretically clean SM prediction $R_K^{\rm SM}=1.0003 \pm 0.0001$~\cite{Bobeth:2007dw} by $2.6\,\sigma$. Combining these anomalies with all other observables for $b\to s \mu^+\mu^-$ transitions, it is found that a scenario with NP in $C_9^{\mu\mu}$ only is preferred compared to the SM by $4.3\,\sigma$ \cite{Altmannshofer:2015sma}.

CMS recently also searched for the decay $h\to\tau\mu$ ~\cite{CMS:2014hha} finding a non-zero result of	${\rm Br} [h\to\mu\tau] = \left( 0.89_{-0.37}^{+0.40} \right)$ which disagrees by about $2.4 \,\sigma$ from 0, i.e. from the SM value. 

Hints for lepton flavour universality violating (LFUV) NP also comes from the BABAR collaboration who performed an analysis of the semileptonic $B$ decays $B\to D^{(*)}\tau\nu$ \cite{BaBar:2012xj}. They find for the ratios ${R}(D^{(*)})\,=\,{\rm Br}(B\to D^{(*)} \tau \nu)/{\rm Br}(B\to D^{(*)} \ell \nu)$:
\begin{eqnarray}
R(D)\,=\,0.440\pm0.058\pm0.042  \,,\qquad R(D^*)\,=\,0.332\pm0.024\pm0.018\,.
\end{eqnarray}
Here the first error is statistical and the second one is systematic. Comparing these measurements to the SM predictions
\begin{eqnarray}
R_{\rm SM}(D)\,=\,0.297\pm0.017 \,, \qquad R_{\rm SM}(D^*) \,=\,0.252\pm0.003 \,,
\end{eqnarray}
we see that there is a discrepancy of 2.2\,$\sigma$ for $\cal{R}(D)$ and 2.7\,$\sigma$ for $\cal{R}(D^*)$ and combining them gives a $3.4\, \sigma$ deviation from the SM~\cite{BaBar:2012xj}. Due to the heavy tau lepton in the final state, these decays are sensitive to charged Higgses \cite{Krawczyk:1987zj}.

\section{New Physics Explanations}

In this section we review NP models which can explain the deviations from the SM discussed in the last section with focus on models with a $Z^\prime$ boson and/or additional Higgs doublets. 

\subsection{Tauonic $B$ Decays}

\begin{figure}[t]
\includegraphics[width=0.49\textwidth]{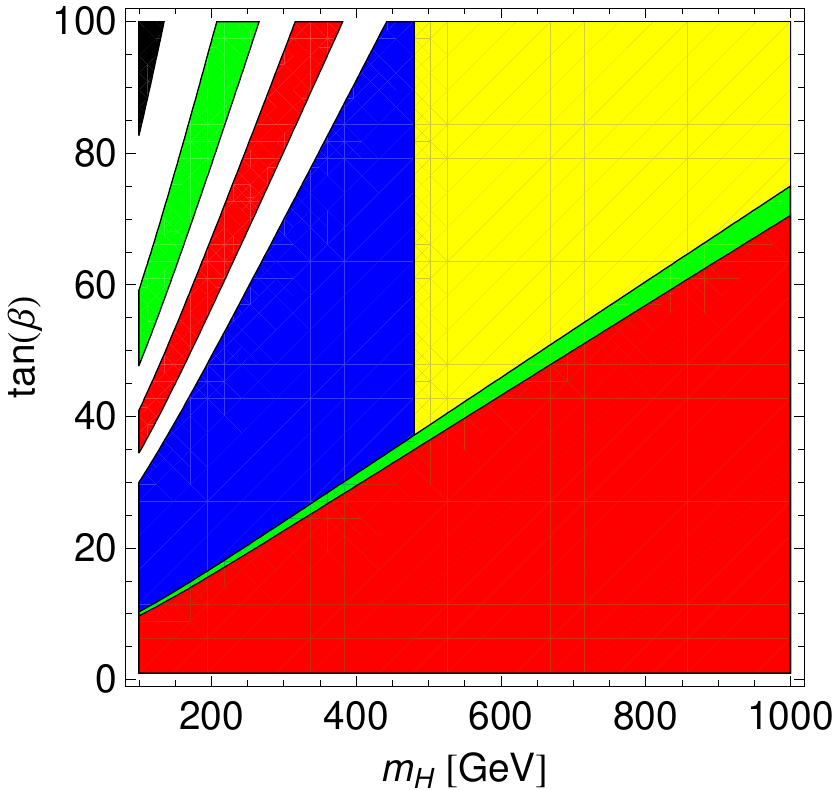}
\includegraphics[width=0.48\textwidth]{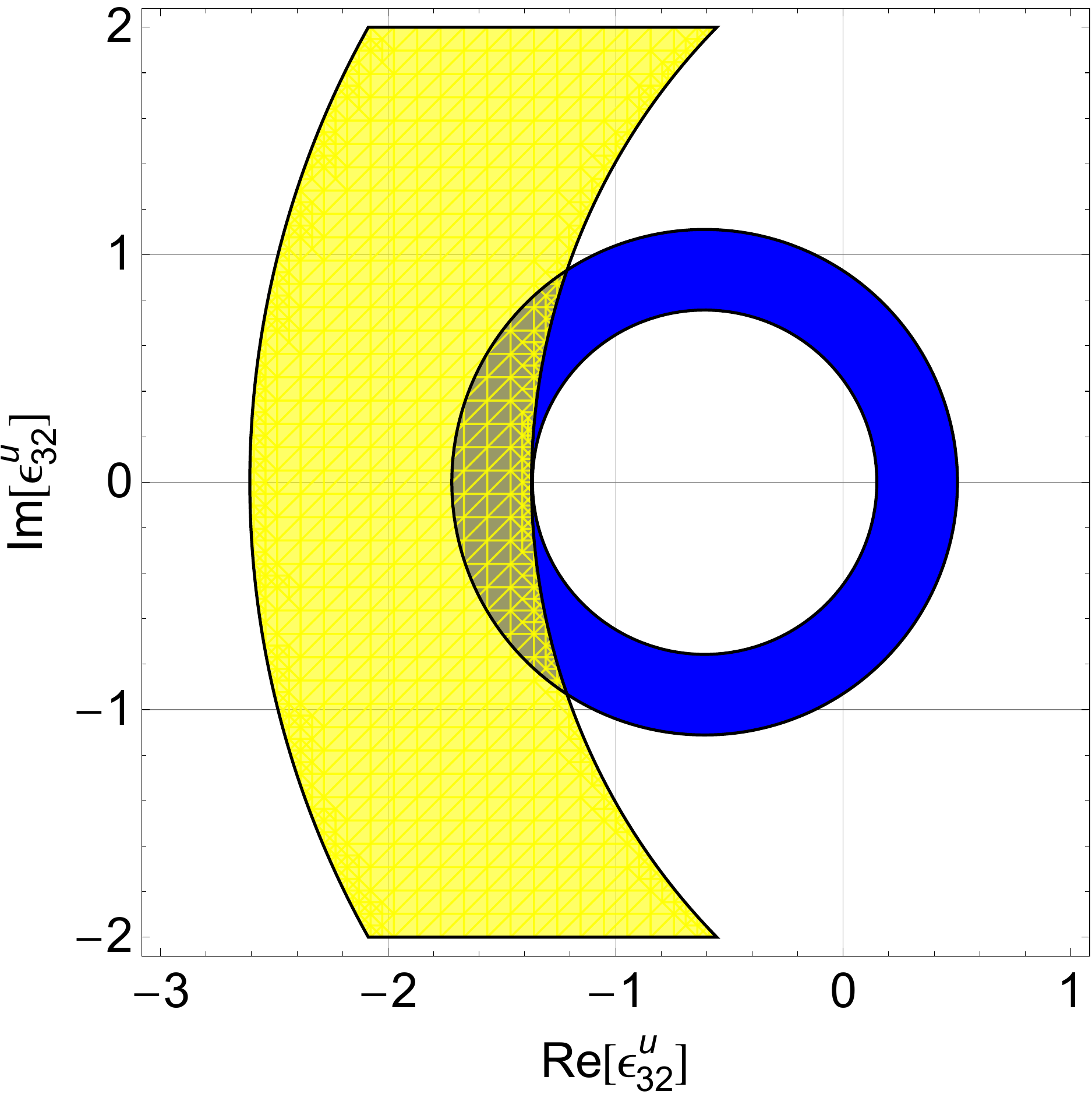}
\caption{
Left: Updated constraints on the 2HDM of type II parameter space. The regions compatible with experiment are shown (the regions are superimposed on each other): $b\to s\gamma$ (yellow) \cite{Misiak:2015xwa}, $B\to D\tau\nu$ (green), $B\to \tau \nu$ (red), $B_{s}\to \mu^{+} \mu^{-}$ (orange), $K\to \mu \nu/\pi\to \mu \nu$ (blue) and $B\to D^*\tau \nu$ (black). Note that no region in parameter space is compatible with all processes since explaining $B\to D^*\tau \nu$ would require very small Higgs masses and large values of $\tan\beta$ which is not compatible with the other observables. To obtain this plot, we added the theoretical uncertainty of the SM linearly on the top of the $2 \, \sigma$ experimental error. Right: Allowed regions in the complex $\epsilon^u_{32}$-plane from $\cal{R}(D)$ (blue) and $\cal{R}(D^*)$ (yellow) for $\tan\beta=40$ and $m_H=800$~GeV.
\label{fig:2HDMII}}
\end{figure}

While a 2HDM of type II (like the MSSM at tree-level) cannot explain the deviations from the SM in tauonic $B$ decays (due to the necessarily destructive interference) without violating bounds from other observables \cite{Crivellin:2013wna} (see left plot in Fig.~\ref{fig:2HDMII}). However, a 2HDM with generic Yukawa coupling (i.e. type III) and large flavour violation in the up-sector can account for $B\to D\tau\nu$ and $B\to D^*\tau\nu$ simultaneously, respecting the constraints from all other observables \cite{Crivellin:2012ye}. For getting the desired effect in $B\to D^{(*)}\tau\nu$, in addition to the parameters already present in the standard type II model ($\tan\beta$ and the heavy Higgs mass $m_H$), only one addition free parameter ($\epsilon^u_{32}$), coupling left-handed top quarks to right-handed charm quarks, is necessary (see right plot in Fig.~\ref{fig:2HDMII}).

\subsection{Anomalies in $b\to s\mu\mu$}

A rather large contribution to operator $(\overline{s}\gamma_\alpha P_L b)(\overline{\mu}\gamma^\alpha \mu)$, as required by the model independent fit~\cite{Altmannshofer:2015sma}, can be achieved in models containing a heavy $Z^\prime$ gauge boson~\cite{Gauld:2013qba,Altmannshofer:2014cfa}. If one aims at explaining also $R(K)$, a contributing to $C_9^{\mu\mu}$ involving muons, but not to $C_9^{ee}$ with electrons is necessary~\cite{Alonso:2014csa,Hiller:2014yaa,Ghosh:2014awa}. This is naturally the case in models with gauged muon minus tauon number ($L_\mu-L_\tau$)~\cite{Altmannshofer:2014cfa,Crivellin:2015mga,Crivellin:2015lwa}\footnote{$Z^\prime$ bosons with the desired couplings can also be obtained in models with extra dimensions \cite{Niehoff:2015bfa}.}. 

In these $Z^\prime$ model the couplings to quarks can be written generically as
\begin{equation}
L\cup g'\left({{{\bar d}}_i}{\gamma ^\mu }{P_L}{{d}_j}{Z'_\mu }\Gamma_{ij}^{{d}L} + {{{\bar d}}}_i{\gamma ^\mu }{P_R}{{d}_j}{Z'_\mu }\Gamma_{ij}^{{d}R}\right)\,.
\end{equation}
where $g^\prime$ is the new $U^\prime(1)$ gauge coupling constant. Unavoidable contributions to \bbs are generated which constrain the coupling to muons to be much larger than the one to $\bar s b$. In the left plot in Fig.~\ref{fig:HiggsPlot} the regions in the $\Gamma^L_{sb}$--$\Gamma^R_{sb}$ plane are shown which are in agreement with \bbs mixing and comply with $b\to s\mu^+\mu^-$ data within $2\,\sigma$. Note that in the symmetry limit $\Gamma^R_{sb}=0$, \bbs mixing puts a upper bound on $\Gamma^L_{sb}$.

\subsection{$h\to \tau\mu$}

LFV SM Higgs couplings are induced by a single operator up to dim-6. Considering only this operator ${\rm Br}[h\to\mu\tau]$ can be up to $10\%$~\cite{Davidson:2012ds}. However, it is in general difficult to get dominant contributions to this operator in a UV complete model, as for example in models with vector-like leptons \cite{Falkowski:2013jya}. Therefore, among the several attempts to explain this $h\to\mu\tau$ observation~\cite{Kopp:2014rva}, most of them are relying on models with extended Higgs sectors. One particularly elegant solution employs a two-Higgs-doublet model (2HDM) with gauged $L_\mu-L_\tau$~\cite{Heeck:2014qea}.

\section{Simultaneous explanation of $b\to s\mu\mu$ and $h\to \tau\mu$ and predictions for $\tau\to3\mu$}

In \cite{Crivellin:2015mga,Crivellin:2015lwa} two models with gauged $L_\mu-L_\tau$ symmetry were presented which can be explain $h\to \tau\mu$ simultaneously with the anomalies in $b\to s\mu\mu$ data (including $R(K)$) giving rise to interesting correlated effects in $\tau\to3\mu$. While in both models the $Z'$ couplings to leptons originate from a gauged $L_\mu-L_\tau$ symmetry, the coupling to quarks is either generated effectively with heavy lepto-quarks charged under $L_\mu-L_\tau$ or directly by assigning horizontal changes to baryons\footnote{For pioneering work on horizontal $U(1)$ gauge symmetries see Ref.~\cite{Terazawa:1976xx}.}.

\subsection{2 Higgs doublets with vector-like quarks}

\begin{figure}
\includegraphics[width=0.45\textwidth]{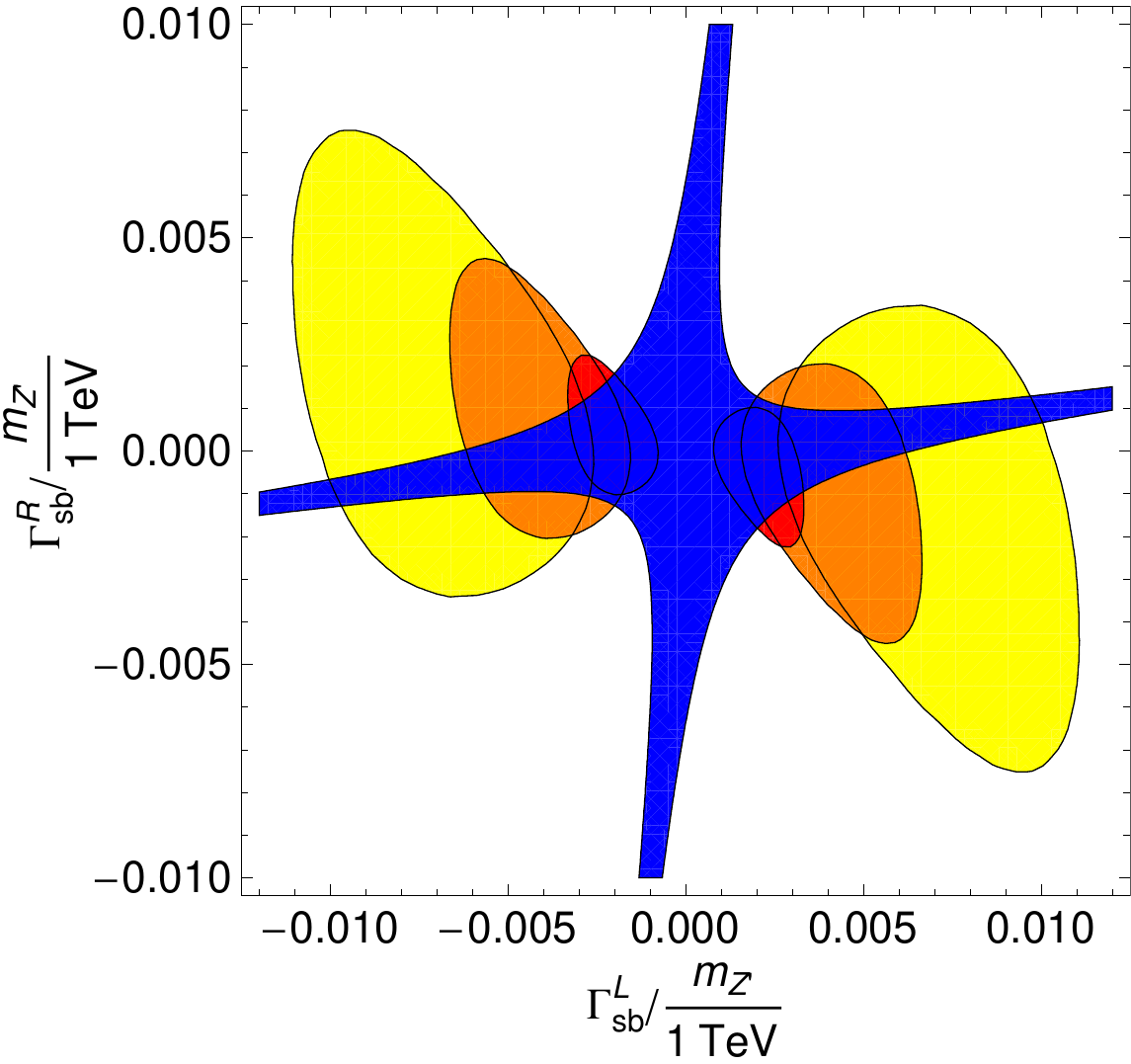}
~~~
\includegraphics[width=0.42\textwidth]{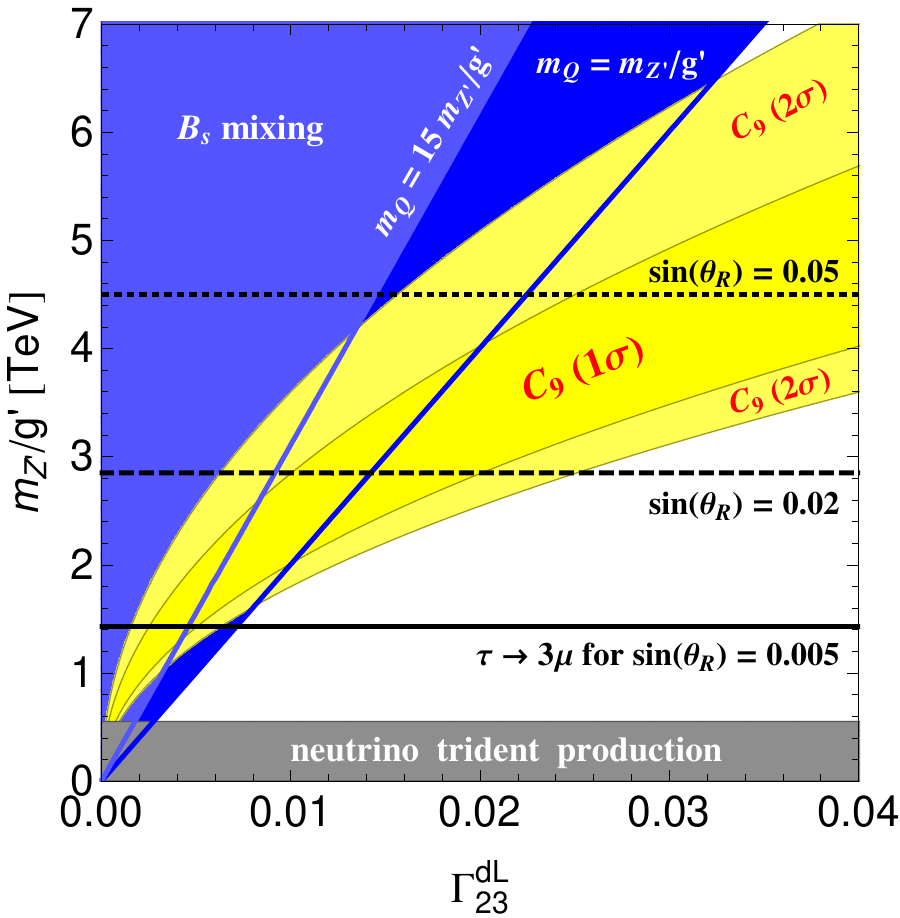}
\caption{ 
Left: Allowed regions in the $\Gamma^{L}_{sb}/M_{Z^\prime}-\Gamma^{R}_{sb}/M_{Z^\prime}$ plane for $g^\prime=1$ from $B_s$-$\overline{B}_s$ mixing (blue), and from the $C^{\mu\mu}_9-C^{(\prime)\mu\mu}_9$ fit of Ref.~\cite{Altmannshofer:2014rta} to $b\to s\mu^+\mu^-$ data, with $\Gamma^V_{\mu\mu}=\pm 1$ (red), $\Gamma_{\mu\mu}^V=\pm 0.5$ (orange) and  $\Gamma^V_{\mu\mu}=\pm 0.3$ (yellow). Note that the allowed regions with positive (negative) $ \Gamma^{L}_{sb}$ correspond to positive (negative) $\Gamma^V_{\mu\mu}$.
Right: Allowed regions in the $\Gamma^{dL}_{23}$--$m_{Z^\prime}/g^\prime$ plane from $b\to s\mu^+\mu^-$ data (yellow) and $B_s$ mixing (blue). For $B_s$ mixing (light) blue 
corresponds to ($m_Q=15 m_{Z^\prime}/g^\prime$) $m_Q=m_{Z^\prime}/g^\prime$. The horizontal lines denote the lower bounds on $m_{Z^\prime}/g^\prime$ from $\tau\to3\mu$ for $\sin(\theta_R)=0.05,\; 0.02,\; 0.005$. The gray region is excluded by NTP.\label{fig:HiggsPlot}}
\end{figure}

Here the model is 2HDM with a gauged $U(1)_{L_\mu-L_\tau}$ symmetry~\cite{Heeck:2014qea}. The $L_\mu-L_\tau$ symmetry is broken spontaneously by the vacuum expectation value of a scalar $\Phi$ (beeing singlet under the SM gauge group) with $Q^{\Phi}_{L_\mu-L_\tau}=1$, leading to the $Z'$ mass $m_{Z'} = \sqrt2 g' \langle\Phi\rangle \equiv g' v_\Phi$. Two Higgs doublets are introduced which break the electroweak symmetry: $\Psi_1$ with $Q^{\Psi_1}_{L_\mu-L_\tau}=-2$ and $\Psi_2$ with $Q^{\Psi_2}_{L_\mu-L_\tau}=0$. Therefore, $\Psi_2$ gives masses to quarks and leptons while $\Psi_1$ couples only off-diagonally to $\tau\mu$:
\begin{align}
\mathcal{L}_Y \ \supset\ &-\overline{\ell}_f Y^\ell_{i}\delta_{fi} \Psi_2 e_i - \xi_{\tau\mu} \overline{\ell}_3 \Psi_1 e_2  -\overline{Q}_f Y^u_{fi} \tilde{\Psi}_2 u_i - \overline{Q}_f Y^d_{fi} \Psi_2 d_i + \mathrm{h.c.}\,.
\label{eq:yukawas}
\end{align}
Here $Q$ ($\ell$) is the left-handed quark (lepton) doublet, $u$ ($e$) is the right-handed up-quark (charged-lepton) and $d$ the right-handed down quark while $i$ and $f$ label the three generations. The scalar potential is the one of a $U(1)$-invariant 2HDM~\cite{Branco:2011iw} with additional couplings to the SM-singlet $\Phi$. We defined as usual $\tan\beta = \langle \Psi_2\rangle/\langle \Psi_1\rangle$ and $\alpha$ is the mixing angle between the neutral CP-even components of $\Psi_1$ and $\Psi_2$ (see for example~\cite{Branco:2011iw}). Therefore, quarks and gauge bosons have standard type-I 2HDM couplings to the scalars. The only deviations from the type I model are in the lepton sector: while the Yukawa couplings $Y^\ell_{i}\delta_{fi}$ of $\Psi_2$ are forced to be diagonal by the ${L_\mu-L_\tau}$ symmetry, $\xi_{\tau\mu}$ gives rise to an off-diagonal entry in the lepton mass matrix:
\begin{equation}
m^\ell_{fi}= \frac{v}{\sqrt{2}}\begin{pmatrix}
y_e\sin\beta &0&0\\
0& y_\mu \sin\beta& 0\\
0&\xi_{\tau\mu} \cos\beta& y_\tau \sin\beta
\end{pmatrix} .
\end{equation}
It is this $\tau$--$\mu$ element that leads to the LFV couplings of $h$ and $Z'$. The mass basis for the charged leptons is obtained by rotations of $(\mu_R,\tau_R)$ and $(\mu_L,\tau_L)$ with the angles $\theta_R$ and $\theta_L$. A non-vanishing angle $\theta_R$ not only gives rise to the LFV decay $h\to\mu\tau$ due to the coupling
\begin{equation}
\frac{m_\tau}{v}\frac{\cos(\alpha-\beta)}{\cos(\beta)\sin(\beta)}\sin(\theta_R)\cos(\theta_R) \bar\tau P_R\mu h\equiv \Gamma^{h}_{\tau\mu}\bar\tau P_R\mu h\,,\label{h0taumu}
\end{equation}
in the Lagrangian, but also leads to off-diagonal $Z'$ couplings to right-handed leptons
\begin{align}
g^\prime Z^\prime_\nu \, (\overline{\mu}, \overline{\tau})
\begin{pmatrix}
 \cos 2\theta_R& \sin 2\theta_R\\
\sin 2\theta_R& - \cos 2\theta_R
\end{pmatrix} \gamma^\nu P_R 
\begin{pmatrix}
 \mu\\
\tau
\end{pmatrix} ,
\end{align}
while the left-handed couplings are to a good approximation flavour conserving. In order to explain the observed anomalies in the $B$ meson decays, a coupling of the $Z'$ to quarks is required as well, not inherently part of $L_\mu-L_\tau$ models (aside from the kinetic $Z$--$Z'$ mixing, which is assumed to be small). Following Ref.~\cite{Altmannshofer:2014cfa}, effective couplings of quarks to the $Z^\prime$  are generated by heavy vector-like quarks \cite{Langacker:2008yv} charged under $L_\mu-L_\tau$. As a result, the couplings of the $Z^\prime$ to quarks are in principle free parameters. In the limit of decoupled vector-like quarks with the quantum numbers of right-handed quarks, only $C_9$ is generated, giving a very good fit to data. The results are shown in the right plot of Fig.~\ref{fig:HiggsPlot} depicting that for small values of $\Gamma^L_{sb}$ and $\theta_R$, $b\to s\mu^+\mu^-$ data can be explained without violating bounds from \bbs mixing or $\tau\to3\mu$. In the left plot of Fig.~\ref{fig:vevplot} the correlations of $b\to s\mu^+\mu^-$ and $h\to\tau\mu$ with $\tau\to3\mu$ are shown, depiciting that consistency with $\tau\to3\mu$ requires large values of $\tan\beta$ (not beeing in conflict with any data as the decoupling limit is a type I model) and future searches for $\tau\to3\mu$ are promising to yield positive results.

\subsection{Horizontal charges for quarks}

\begin{figure}[t]
\includegraphics[width=0.41\textwidth]{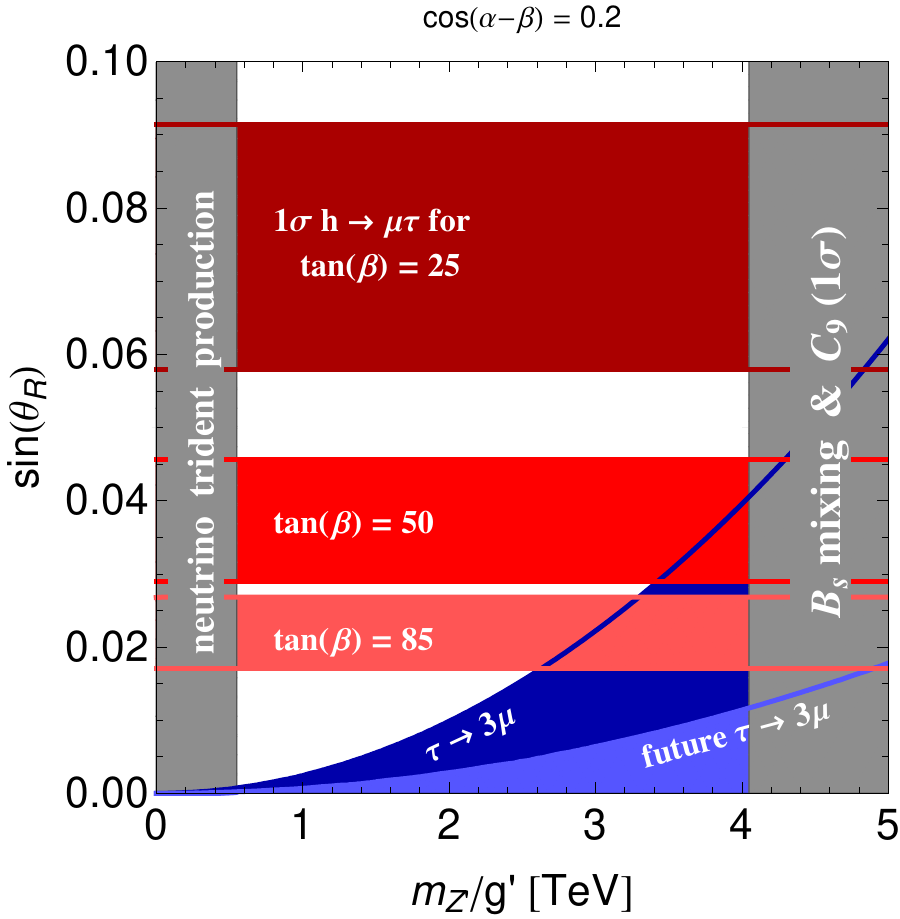}
\includegraphics[width=0.59\textwidth]{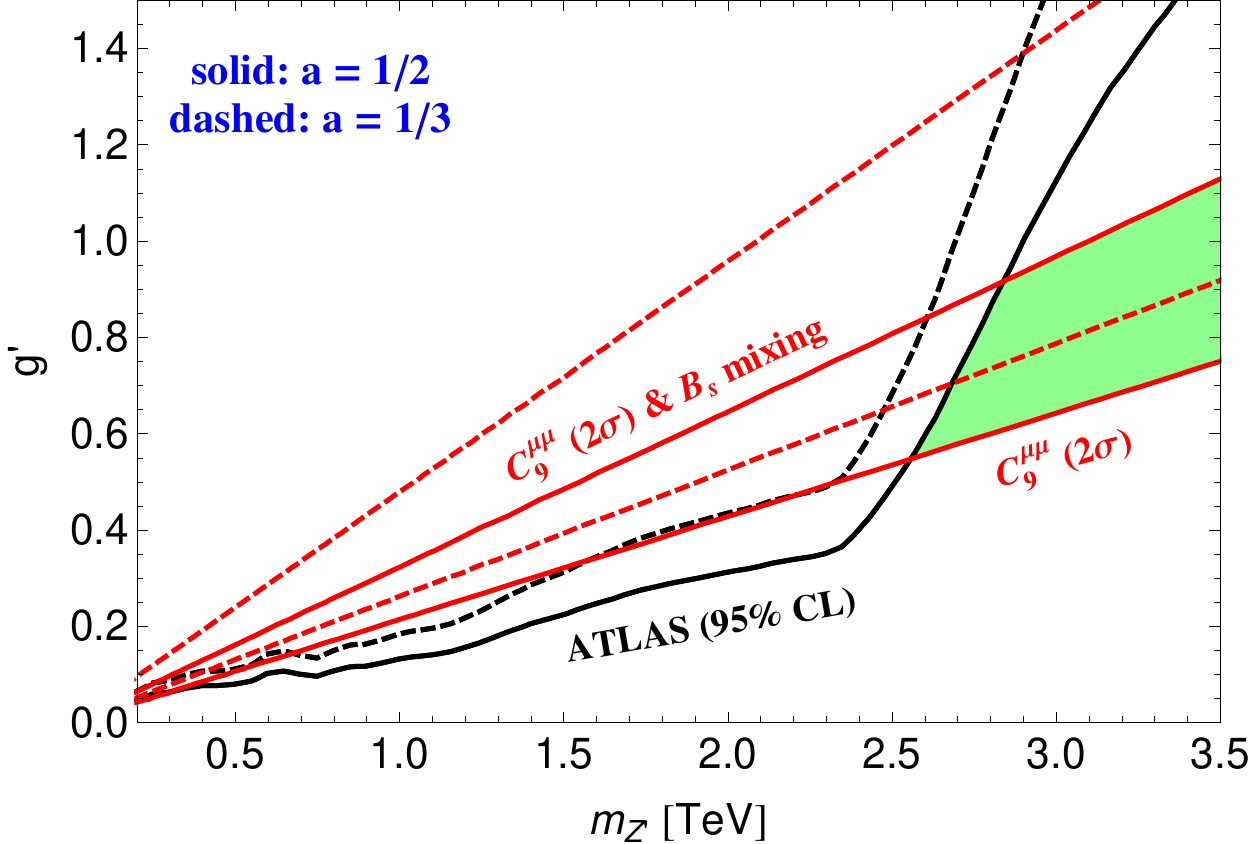}
\caption{Left: Allowed regions in the $m_{Z'}/g'$--$\sin (\theta_R)$ plane: the horizontal stripes correspond to $h\to\mu\tau$ ($1\sigma$) for $\tan\beta=85,\,50,\,25$ and $\cos (\alpha-\beta)=0.2$, (light) blue stands for (future) $\tau\to 3\mu$ limits at $90\%$~C.L. The gray regions are excluded by NTP or $B_s$--$\overline{B}_s$ mixing in combination with the $1\,\sigma$ range for $C_9$. \newline
Right: Limits on $q\overline{q}\to Z' \to \mu\overline{\mu}$ from ATLAS~\cite{Aad:2014cka} (black, allowed region down right) and the $2\sigma$ limits on $C_9^{\mu\mu}$ to accommodate $b\to s\mu^+\mu^-$ data (red, allowed regions inside the cone). Solid (dashed) lines are for $a=1/2$ ($a=1/3$). For $a =1/2$, the green shaded region is allowed (similar for $a= 1/3$ using the dashed bounds).}
\label{fig:vevplot}
\end{figure}

In order to avoid the introduction of vector-like quarks, one can introduce flavour-dependent charges to quarks as well \cite{Crivellin:2015lwa}. Here, the first two generations should have the same charges in order to avoid very large effects in $K$--$\overline{K}$ or $D$--$\overline{D}$ mixing, generated otherwise unavoidably due to the breaking of the symmetry necessary to generate the measured Cabibbo angle of the CKM matrix. If we require in addition the absence of anomalies, we arrive at the following charge assignment for baryons	$Q'(B)= (-a,\,-a,\,2a )$.
Here $a \in {\cal Q}$ is a free parameter of the model with important phenomenological implications. 
In this model, at least one additional Higgs doublet which breaks the flavour symmetry in the quark sector is required, and one more is needed if one attempts to explain $h\to\tau\mu$. In case the mixing among the doublets is small, the correlations among $h\to\tau\mu$, $b\to s\mu^+\mu^-$ and $\tau\to 3\mu$ are the same is in the model with vector-like quarks discussed in the last subsection and shown in the left plot of Fig.~\ref{fig:vevplot}.

The low-energy phenomenology is rather similar to the one of the model with vector like quarks considered in the last section, but the contributions to $B_s-\overline{B}_s$ mixing are directly correlated to $B_d-\overline{B}_d$ and $K-\overline{K}$ mixing as all flavour violation is due to CKM factor. However, concerning direct LHC searches, the implications are very different, as the $Z^\prime$ boson can be directly produced on-shell as a resonance in $p\bar p$ collisions since it couples to quarks of the first generation. The resulting strong bounds are shown in right plot of Fig.~\ref{fig:vevplot} where they are compared to the allowed regions from $B_s-\overline{B}_s$ mixing and $b\to s\mu^+\mu^-$ data for different values of $a$.

\subsection{Lepton flavour violating $B$ decays}

As lepton flavour universality is violated in $R(K)$ and $B\to D^{(*)}\tau\nu$, and $h\to\tau\mu$ even violates lepton flavour, it is interesting to examine the possibility of observing lepton flavour violating $B$ decays \cite{Glashow:2014iga}. Here we review $B\to K^{(*)}\tau^\pm\mu^\mp$ and $B_s\to \tau^\pm\mu^\mp$ in $Z^\prime$ models with generic couplings to fermions \cite{Crivellin:2015era}. While in the UV complete model of Refs.~\cite{Crivellin:2015mga,Crivellin:2015lwa} the branching ratios for LFV $B$ decays are tiny, in general these processes are proportional to $\Gamma_{\mu\tau}\Gamma_{sb}$ and can be large in the presence of sizable flavour violation in the quark and in the lepton sector. As we can see from the left plot in Fig.~\ref{fig:HiggsPlot}, $\Gamma^L_{sb}$ can only be large if there are cancellations originating from $\Gamma^R_{sb}$ having the same sign but being much smaller. Therefore, the branching ratios for LFV $B$ decays are bounded by fine tuning together with $\tau\to3\mu$ and $\tau\to\mu\nu\nu$ limiting $\Gamma_{\mu\tau}$. As a result, we find in a scenario in which NP contributions to $C_9^{\ell\ell^\prime}$ only are generated
\begin{align}
{\rm Br}\left[B\to K^{(*)}\tau^\pm\mu^\mp\right]&\le 2.2(4.4)\times 10^{-8}(1+X_{B_s})\,,\\
{\rm Br}\left[B_s\to\tau^\pm\mu^\mp\right]&\le 4.3\times 10^{-8}(1+X_{B_s})\,,
\end{align}
where $X_{B_s}$ measures the degree of fine tuning in the $B_s$ system. Note that these limits are obtained for $\Gamma_{\mu\mu}=0$ (which corresponds to $C_9^{\mu\mu}=0$) and are even stronger for non-vanishing values of $\Gamma_{\mu\mu}$. For $\mu e$ final states the possible branching ratios are much smaller due to the stringent constraints from $\mu\to e\gamma$ and $\mu\to e\nu\nu$.

\section{Conclusion}

In these proceedings we reviewed the impact of the indirect hint for physics beyond the SM in the flavour sector obtained by BABAR, LHCb and CMS on models of NP with focus on models with $Z^\prime$ bosons and/or additional Higgs doublets. While a prime candidate for the explanation of the anomalous $b\to s\mu^+\mu^-$ data is a $Z^\prime$ boson, $h\to\tau\mu$ as well as $B\to D^{(*)}\tau\nu$ can be most naturally explained by an extended scalar sector. Interestingly, models with gauged $L_\mu-L_\tau$ can explain $b\to s\mu^+\mu^-$ data and $h\to\tau\mu$ simultaneously, predicting sizable branching ratios of $\tau\to3\mu$, potentially observable in future experiments. While the UV complete models \cite{Crivellin:2015mga,Crivellin:2015lwa} predict tiny branching ratios for LFV $B$ decays, these decays can have sizable branching fractions for $\tau\mu$ final states in generic $Z^\prime$ models in the presence of significant fine-tuning in the $B_s-\overline{B}_s$ system.  

\acknowledgments

I thank the organizers, especially Gino Isidori, for the invitation to "Les Rencontres de Physique de la Valle d'Aoste" and the opportunity to present these results. I also thank Stefan Pokorski and Giancarlo D'Ambrosio for proofreading. This work is supported by a Marie Curie Intra-European Fellowship of the European Community's 7th Framework Programme under contract number (PIEF-GA-2012-326948).

\end{document}